\documentclass{article}
\usepackage{ijcai26}

\usepackage{times}
\usepackage{soul}
\usepackage{url}
\usepackage[hidelinks]{hyperref}
\usepackage[utf8]{inputenc}
\usepackage[small]{caption}
\usepackage{graphicx}
\usepackage{amsmath}
\usepackage{amsthm}
\usepackage{booktabs}
\usepackage{algorithm}
\usepackage{algorithmic}
\usepackage[switch]{lineno}
\usepackage{booktabs}   % 专业三线表
\usepackage{tabularx}   % 自动换行
\usepackage{multirow}   % 跨行合并
\usepackage[table]{xcolor} % 颜色支持
\usepackage{makecell}   % 单元格内辅助换行
\usepackage{xltabular} % 核心：支持跨页和自适应宽度
\usepackage{array}
\usepackage{tabularx}
\usepackage{makecell}
\usepackage{CJKutf8}
\usepackage{stfloats}

\title{Evaluation of Large Language Models in Legal Applications: \\ Challenges, Methods, and Future Directions}

\author{
Yiran Hu$^{1,2,3}$\thanks{Equal contribution.}
\and
Huanghai Liu$^{1*}$
\and
Chong Wang$^{1*}$
\and
Kunran Li$^1$\and
Tien-Hsuan Wu$^2$\and
Haitao Li$^1$\and
Xinran Xu$^4$\and
Siqing Huo$^3$\and
Weihang Su$^1$\and
Ning Zheng$^1$\and
Siyuan Zheng$^4$\and
Qingyao Ai$^1$\and
Yun Liu$^1$\and
Renqun Bian$^5$\and
Yiqun Liu$^1$\and
Charles L.A. Clarke$^3$\and
Weixing Shen$^1$\And
Ben Kao$^2$\thanks{Corresponding author.}
\\
\affiliations
$^1$Tsinghua University\
$^2$The University of Hong Kong\
$^3$University of Waterloo\\
$^4$Shanghai Jiaotong University\
$^5$Peking University\\
\emails
huyr17@outlook.com,
liuhh23@mails.tsinghua.edu.cn
}
\begin{document}

\maketitle

\begin{abstract}
% Large language models (LLMs) are increasingly deployed across a wide range of legal applications, including judicial decision support, legal practice assistance, and public-facing legal services. While these models demonstrate promising capabilities in handling legal knowledge and reasoning tasks, their integration into real-world legal workflows raises significant challenges related to reliability, reasoning soundness, fairness, and trustworthiness. As a result, systematically evaluating LLM performance in legal tasks has become a critical prerequisite for their responsible adoption. This survey adopts an application-driven perspective, starting from how LLMs are used in real judicial contexts, including applications for judges, lawyers, and the general public. We analyze the challenges that arise when LLMs operate in legal practice, such as reasoning reliability, procedural correctness, fairness, and trustworthiness. Using these challenges as a guiding thread, we review and categorize existing evaluation methods and benchmarks for LLMs in legal tasks, analyzing their task design, datasets, and evaluation metrics. We further discuss whether current methods adequately address the challenges posed by real-world legal applications. Finally, we identify key limitations in existing evaluation approaches and outline future research directions toward more realistic, reliable, and application-oriented evaluation frameworks for LLMs in legal domains.

Large language models (LLMs) are being increasingly integrated into legal applications, including judicial decision support, legal practice assistance, and public-facing legal services. While LLMs show strong potential in handling legal knowledge and tasks, their deployment in real-world legal settings raises critical concerns beyond surface-level accuracy, involving the soundness of legal reasoning processes and trustworthy issues such as fairness and reliability. Systematic evaluation of LLM performance in legal tasks has therefore become essential for their responsible adoption. This survey identifies key challenges in evaluating LLMs for legal tasks grounded in real-world legal practice. We analyze the major difficulties involved in assessing LLM performance in the legal domain, including outcome correctness, reasoning reliability, and trustworthiness. Building on these challenges, we review and categorize existing evaluation methods and benchmarks according to their task design, datasets, and evaluation metrics. We further discuss the extent to which current approaches address these challenges, highlight their limitations, and outline future research directions toward more realistic, reliable, and legally grounded evaluation frameworks for LLMs in legal domains.

% This survey adopts an application-driven perspective, examining how LLMs are used by judges, lawyers, and the general public. We analyze the challenges that arise in practical legal contexts and review existing evaluation methods and benchmarks accordingly, categorizing them by task design, datasets, and evaluation metrics. We further assess whether current approaches adequately address real-world legal risks and limitations, and outline future research directions toward more realistic, reliable, and application-oriented evaluation frameworks for LLMs in legal domains.
\end{abstract}

\section{Introduction}

As large language models (LLMs) continue to advance in capability, they are increasingly being applied to the legal domain~\cite{sun2024lawluo,li2024legalagentbenchevaluatingllmagents}. Typical applications include assisting laypersons in understanding legal issues, supporting lawyers in legal practice, and providing auxiliary assistance to judges in judicial decision-making~\cite{su-etal-2024-stard,li2024legalagentbenchevaluatingllmagents,gao2024enhancinglegalcaseretrieval,chen2024agentcourtsimulatingcourtadversarial}. This raises a fundamental question: \textbf{Are LLMs truly qualified to enter the legal domain, and which types of legal tasks can they reliably perform?}

Previous studies~\cite{katz2024gpt,freitas2023does} have shown that LLMs are able to pass legal examinations, and some works~\cite{xiao2018cail2018,lyu2023multi} further demonstrate that LLMs can achieve high performance in legal judgment prediction tasks. However, due to the intrinsic complexity of real-world legal scenarios and the demanding nature of legal reasoning, evaluating LLMs solely based on standardized exam-style questions or prediction accuracy is insufficient to comprehensively assess their performance in real legal applications. Consequently, a growing number of benchmarks~\cite{fei2023lawbench,legalbench,li2024lexeval} have been proposed to evaluate LLMs from multiple perspectives in legal tasks.
% The objective of evaluating LLMs in the legal domain is to \textbf{systematically and automatically assess their capabilities across multiple dimensions in a comprehensive manner.}

% However, evaluating LLMs in legal tasks is significantly more challenging than conducting general-purpose evaluations. Addressing legal problems requires strong logical reasoning abilities. Even if a model produces a correct final answer for a legal task, flawed reasoning processes undermine its ability to meaningfully apply legal knowledge to analyze and resolve legal issues. Although several studies have attempted to evaluate the legal reasoning capabilities of LLMs, existing approaches still fall short of providing fine-grained, accurate, and realistic assessments of legal reasoning.

% Another major challenge lies in the trustworthiness evaluation of LLMs in the legal domain. Legal decisions are closely tied to individuals' lives and social interests, and any legal error may have serious consequences for multiple stakeholders. Therefore, ensuring that models are fair, reliable, and safe is a prerequisite for deploying LLMs in legal applications. While prior work has explored the trustworthiness of LLMs in legal tasks, most studies focus on isolated aspects of trustworthiness and lack systematic, large-scale, and foundational evaluations.

% In summary, existing benchmarks for evaluating LLMs in the legal domain primarily focus on three dimensions:
% \textbf{(1) output accuracy, (2) trustworthiness, and (3) logical ability.}
Evaluating LLMs in legal tasks is significantly more challenging than conducting general-purpose evaluations~\cite{huang2023c}. Existing studies~\cite{hu2025llms,liu2024judges} show that LLMs are increasingly being applied in real-world court settings. In real legal applications, LLMs face many issues beyond whether an answer is correct. Evaluation should not focus solely on the final result, but also consider the reasoning process and system-level constraints. The aim of evaluating LLMs in the legal domain is to \textbf{systematically and automatically assess their capabilities across multiple dimensions,} reflecting not only whether models arrive at correct outcomes, but also how those results are produced and under what constraints they remain valid.

 From the process perspective, addressing legal problems requires strong logical reasoning abilities; a model may produce a legally correct answer while relying on flawed or invalid reasoning steps. For example, when an LLM serves as a judicial assistant, a key factor in whether the public can be convinced lies in its reasoning process, such as whether it cites authentic and correct legal provisions and whether it performs logically sound inferences based on both legal rules and factual circumstances. From the constraint perspective, LLMs need to handle legal cases fairly and safely. For instance, an LLM should not exhibit discrimination based on gender or geographic origin when making judgments. Given that legal decisions directly affect individuals' rights and social welfare, ensuring fairness, reliability, and safety is a prerequisite for real-world adoption~\cite{wang2023decodingtrust}. Legal applications operate under strict normative and social constraints: even when both the final result and the reasoning appear correct, a model may still be unsuitable for deployment if it exhibits unfair bias, lacks robustness, or fails under adversarial or distributional shifts.

% Evaluating LLMs in legal tasks is significantly more challenging than conducting general-purpose evaluations\cite{huang2023c}. Addressing legal problems requires strong logical reasoning abilities, a model may produce a legally correct answer while relying on flawed or invalid reasoning steps. Such cases reveal limitations in the model's ability to meaningfully apply legal principles, even when surface-level accuracy appears high. Although prior studies\cite{legalbench} have attempted to assess legal reasoning, existing methods often lack fine-grained, realistic analysis of whether reasoning aligns with accepted legal logic and doctrines.

% Beyond outcomes and reasoning processes, a third challenge arises from trustworthiness. Legal applications operate under strict normative and social constraints: even when both the final result and reasoning appear correct, a model may still be unsuitable for deployment if it exhibits unfair bias, lacks robustness, or fails under adversarial or distributional shifts. Given that legal decisions directly affect individuals' rights and social welfare, ensuring fairness, reliability, and safety is a prerequisite for real-world adoption~\cite{wang2023decodingtrust}. However, existing work on trustworthiness typically examines isolated aspects, rather than evaluating system-level behavior in a comprehensive and principled manner.

% In summary, existing benchmarks for evaluating LLMs in the legal domain can be conceptually organized into three overarching dimensions: \textbf{Outcome correctness, Trustworthiness, Legal Reasoning}

In summary, when LLMs are applied to real-world legal settings, they need to be evaluated across the entire pipeline (Result, Process, Constraint). These three stages correspond to three evaluation dimensions: \textbf{Output Accuracy (Result), Legal Reasoning (Process), and Trustworthiness (Constraint)}. Only through effective evaluation along these three dimensions can models be applied more reliably in practice. Based on these three challenges, we categorize existing benchmarks and select representative benchmarks for analysis.

\begin{figure*}[t]
    \centering
    \includegraphics[width=0.8\textwidth, height=0.3\textheight]{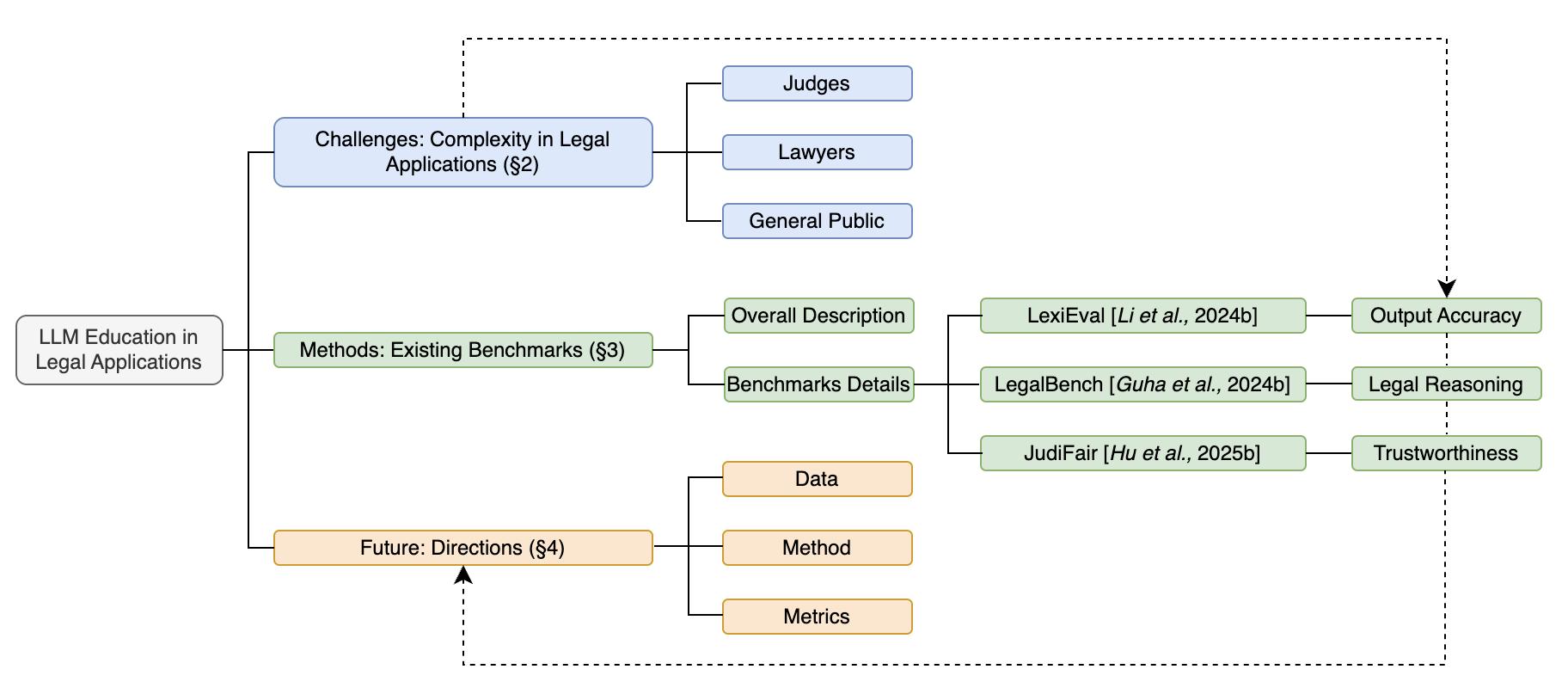}

    \caption{Overview of the proposed framework.}

    \label{fig:overview}
\end{figure*}

% \textbf{(1) Result-oriented evaluation}, which focuses on outcome correctness; \textbf{(2) Process-oriented evaluation}, which examines the validity and soundness of legal reasoning; and \textbf{(3) Constraint-oriented evaluation}, which assesses trustworthiness, including fairness, robustness, and safety. This tripartite perspective highlights both the progress made and the critical gaps that remain in current evaluation practices.

\textbf{Outcome Accuracy} aims to assess a model's legal knowledge and its overall task performance. Existing datasets are mainly constructed around legal examinations~\cite{li2024lexeval,fan2025lexam} and judicial decision prediction tasks~\cite{xiao2018cail2018}, with evaluation metrics such as accuracy for multiple-choice questions, as well as ROUGE and BERT-based similarity measures for open-ended legal responses. \textbf{Trustworthiness} evaluates the potential risks introduced by LLMs when performing legal tasks~\cite{wang2023decodingtrust}. Current studies~\cite{hu2025llms,hu2025jandh} mainly investigate ethical concerns, fairness, and robustness, often using counterfactual analysis and adversarial attacks. \textbf{Legal Reasoning} measures whether the reasoning process employed by the model aligns with legal reasoning principles. Existing work~\cite{akyurek2025prbench} typically relies on fine-grained legal reasoning rubrics annotated by human experts to score the model's reasoning steps.

% Based on evaluations using these benchmarks, we observe that while current LLMs may achieve strong performance on relatively simple legal question-answering tasks (e.g., legal exams), they still exhibit significant limitations in terms of trustworthiness and logical reasoning. These findings provide important guidance for improving the deployment of LLMs in real-world legal scenarios.

Despite the progress made by existing research, many challenges remain unresolved. For instance, trustworthiness evaluations still lack coverage of critical dimensions such as privacy risks and toxicity in legal contexts. Evaluations of logical inference heavily rely on manually annotated rubrics, making them difficult to scale to large datasets. Moreover, certain task-specific issues, like legal hallucination, have not yet been systematically evaluated. These limitations indicate that current evaluation frameworks for LLMs in legal tasks are far from complete, leaving substantial room for further exploration.

As Figure \ref{fig:overview} shows, in this survey, we summarize existing studies on evaluating LLMs in the legal domain, identify key challenges, and outline promising future research directions.

The contributions of this survey are threefold:

\begin{enumerate}
\item We identify major challenges in applying LLMs to the legal domain and discuss potential solutions.
\item We systematically summarize and categorize existing evaluation methods and benchmarks for legal LLMs. For details, see \url{https://github.com/THUYRan/Evaluation-of-LLMs-in-Legal-Applications}
\item We align current evaluation approaches with outstanding challenges, highlighting open problems and future research directions.
\end{enumerate}

\section{Challenges: Risks and Complexity of LLM Applications in Law}

A comprehensive, scientific, and reliable evaluation of LLM performance in legal scenarios needs to start from their specific applications. A prerequisite for meaningful LLM evaluation is to understand where and how LLMs are being deployed in legal workflows, and how these deployments affect legal decision-making and everyday life. By systematically examining the legal tasks to which LLMs are applied, researchers can better identify the unique challenges involved in evaluating LLMs within the legal domain.

In this section, we review representative applications of LLMs across three user groups — judges, lawyers, and the general public. Although these applications differ in function, authority, and risk profile, they illustrate the breadth of LLM adoption in legal contexts and highlight why legal AI evaluation needs to be carried out across the entire pipeline.

\subsection{Applications to Judges}

LLMs are increasingly incorporated into judicial workflows, supporting tasks such as \textbf{case triage, legal document drafting, and decision review}~\cite{alon2023human}. Judicial applications place LLMs directly within the exercise of legal authority, imposing especially stringent requirements on legal accuracy, reasoning quality, and ethical compliance. Their growing adoption in judicial contexts thus underscores the need for rigorous and multidimensional evaluation.

One prominent application is case classification and workflow triage at the filing stage. LLM-based systems can assess case complexity and recommend appropriate procedures, improving efficiency and reducing administrative burdens. However, misclassification or incorrect issue identification may result in improper procedural routing, potentially affecting parties' substantive rights. Evaluation in this context must therefore go beyond output accuracy to assess fact sensitivity and procedural reasoning.

% LLMs are also used to assist judicial reasoning and document preparation, including verdict consistency checks and draft generation. While such systems may help identify inconsistencies with precedents or statutory norms, they introduce deeper evaluation challenges. Even legally plausible outputs may rely on flawed reasoning, selective precedent use\cite{alon2023human}, or insufficient justification, which can undermine fairness and transparency. Given the high societal impact of judicial decisions, evaluation must jointly assess legal correctness, reasoning soundness, transparency, and trustworthiness.

Judicial settings further amplify the importance of trustworthy evaluation. Judicial decisions carry profound social consequences, and errors or biases introduced by LLMs may directly erode public trust in the legal system. When LLMs are used to assist judicial reasoning and document preparation, including verdict consistency checks and draft generation, risks such as selective adherence to AI recommendations, limited accountability, and insufficient transparency expose the limitations of existing evaluation approaches that focus on isolated performance metrics. These challenges underscore that evaluating LLMs for judicial applications is both technically demanding and socially critical.

\subsection{Applications to Lawyers}

LLMs have rapidly become integral tools for legal professionals, assisting with tasks such as \textbf{legal research, contract review\footnote{\url{https://law.co/contract-review-automation}}, document drafting, case summarization, and evidence analysis}~\cite{perlman2023implications,tan2023chatgpt,Noonan2023CreativeMA}. By automating routine work, LLMs enable lawyers to focus on higher-value strategic tasks~\cite{tool2023lawyers,lawfirm2023lawyers}, which has driven widespread adoption in legal practice.

However, even common applications, such as precedent retrieval or document summarization, pose significant evaluation challenges. Hallucinated citations, jurisdictional mismatches, incomplete legal coverage, and subtle terminological errors may lead to misleading outputs. As a result, evaluating LLMs for legal assistance requires assessing reliability, factual grounding, and jurisdictional awareness, rather than relying solely on surface-level accuracy.

In litigation-related contexts, LLMs are explored for argument preparation, evidence analysis, and simulated court interactions. These applications further highlight limitations in current evaluation approaches. Although models may generate persuasive arguments, they often fail to capture procedural dynamics, ethical constraints, and adversarial context. Similarly, in non-litigation settings such as due diligence, compliance, and intellectual property analysis, errors may carry substantial legal and commercial risks~\cite{howlett2023chatgpt,murray2023artificial}. These scenarios emphasize that evaluation must extend beyond task performance to include explainability, robustness, and alignment with professional norms.

\subsection{Applications to the General Public}

LLM-based legal tools offer new opportunities to improve public access to legal information. These tools can \textbf{explain statutes, summarize judgments, retrieve relevant policies\footnote{\url{https://tongyi.aliyun.com/farui}}}, \textbf{and guide users through basic legal procedures}, reducing barriers posed by legal costs~\cite{mitchell2025lawyerconsult}, technical language~\cite{whalen2015judicial,martinez2024even}, and procedural complexity.

At the same time, public-facing applications introduce particularly severe evaluation challenges. Public users often lack the expertise to verify legal advice, amplifying the risks of hallucination, outdated information, or jurisdictional misapplication. Moreover, users frequently provide vague or incomplete descriptions, requiring LLMs to infer intent, identify missing information, and manage uncertainty. Evaluation must therefore consider robustness to noisy input, contextual understanding, and the model's ability to recognize and communicate limitations.

Because public users may rely heavily on system outputs without independent verification, reliability and safety become paramount. While current LLMs are effective for preliminary legal guidance, they remain insufficiently reliable for complex or high-stakes decision-making. This gap highlights the necessity of comprehensive evaluation frameworks capable of determining whether LLM-based legal tools can be responsibly deployed for public-facing legal assistance.

% \vspace{-0.6cm}
\section{Methods: Evaluation Benchmarks for LLMs in the Legal Domain}

As Table \ref{tab:legal_benchmarks_v2} shows, based on the different tasks that LLMs perform in real-world legal settings, many datasets have been proposed to evaluate LLMs. These evaluations are mostly conducted from the perspective of a single task and lack systematic, comprehensive, multi-dimensional assessment. However, these benchmarks provide a foundation for future multi-task evaluation. Subsequent work can build on these single-task evaluations and further integrate them.

In this section, we start from these single-task evaluations and categorize them into generation tasks and decision tasks. We introduce the tasks, datasets, and evaluation metrics for each category. We then present multi-task benchmarks, selecting one of the most well-known benchmarks for each of the three dimensions — output accuracy, legal reasoning, and trustworthiness — for detailed discussion, followed by an overall analysis and summary of evaluation along each dimension.

\subsection{Single-Task Benchmarks}
The evaluation of legal LLMs is typically structured around the characteristics of downstream tasks~\cite{li2023muser,hu2025jandh,ma2021lecard,chalkidis2017extracting,yao2022leven}. We categorize these benchmark tasks into two groups, Generation Tasks and Decision Tasks.
% as illustrated in Table~\ref{tab:evaluation-metrics}.
% In the table, we show example tasks, evaluation methods, and quantitative metrics. 

% \begin{table*}[h]
% \centering
% \caption{Benchmark Tasks for Legal LLM Evaluation.}
% \label{tab:evaluation-metrics}
% \small
% \renewcommand\arraystretch{1.4}

% \begin{tabular}{|m{3cm}|m{6cm}|m{6cm}|}
% \hline
%  & \textbf{Generation Tasks} & \textbf{Decision Tasks} \\ \hline

% \textbf{Examples}
% & Document summarization, fact extraction, legal reasoning, dialogue generation.
% & Law exam multiple-choice questions, case retrieval, opinion alignment. \\ \hline

% \textbf{Evaluation Criteria}
% & Semantic similarity and relevance to reference answers.
% & Classifier, recommender, and ranker performance. \\ \hline

% \textbf{Common Metrics}
% & ROUGE, BERTScore, LLM-as-a-Judge.
% & Precision, Recall, F1, Micro-F1, EM, NDCG@k, MRR. \\ \hline

% \end{tabular}
% \end{table*}

\urlstyle{same}

{\bf Generation Tasks}. These tasks~\cite{shen2022multi,louis2024interpretable,kornilova2019billsum} require a legal LLM to generate text. For example, a legal LLM can summarize a lengthy court opinion into its key facts, arguments, reasoning, and outcomes. To evaluate a legal LLM on such tasks, instances of these tasks along with their reference answers are needed. The model's performance is assessed based on how closely its generated output resembles the reference answer in terms of their semantic similarity and relevance. Common quantitative metrics used include ROUGE-L~\cite{steffes2023evaluating,mullick2022evaluation} and BERT-Score~\cite{kumar2024large,benedetto2023benchmarking,joshi2024tur}. Alternatively, the generated output can be assessed by human assessors for a more comprehensive and qualitative evaluation. With advancements in LLMs, {\it LLM-as-a-judge}, which is a method that uses an LLM to simulate human evaluators in assessing the quality of other models' output, is gaining interest as an automated assessment tool~\cite{cui2023chatlaw,li2024llms}. For example, the CAIL-2024 (Challenge of AI in Law) competition introduced a subjective evaluation metric in the legal consultation track\footnote{\url{http://cail.cipsc.org.cn/task_summit.html?raceID=4&cail_tag=2024}}, which involves simulated scoring by LLMs. This metric evaluates the coherence of generated dialogues, the relevancy of answers, the accuracy of legal knowledge, and the correctness of legal references and provisions. 

% {\bf Generation Tasks}. These tasks require a legal LLM to generate text. For example, a legal LLM can summarize a lengthy court opinion into its key facts, arguments, reasoning, and outcomes. To evaluate a legal LLM on such tasks, instances of these tasks along with their reference answers are needed. The model's performance is assessed based on how closely its generated output resembles the reference answer in terms of their semantic similarity and relevance. Common quantitative metrics used include ROUGE-L and BERT-Score. Alternatively, the generated output can be assessed by human assessors for a more comprehensive and qualitative evaluation. With advancements in LLMs, {\it LLM-as-a-judge}, which is a method that uses an LLM to simulate human evaluators in assessing the quality of other models' output, is gaining interests as an automated assessment tool. For example, the CAIL-2024 competition introduced a subjective evaluation metric in the legal consultation track\footnote{\url{http://cail.cipsc.org.cn/task_summit.html?raceID=4&cail_tag=2024}}, which involves simulated scoring by LLMs. This metric evaluates the coherence of generated dialogues, relevancy of answers, accuracy of legal knowledge, and the correctness of legal references and provisions. 

{\bf Decision Tasks}.
These tasks require a legal LLM to make decisions, such as choosing or ranking options from a set of alternatives, or determining suitable outcomes. Typical tasks include classification (e.g., in legal multiple-choice question-answering~\cite{pahilajani2024nlp}, the LLM needs to classify whether an answer is correct for a given question), extraction and ranking (e.g., in legal case retrieval~\cite{feng2024legal,padiu2024extent}, the LLM is required to identify relevant court judgments and rank them based on their relevance), and prediction~\cite{wu2023precedent,wang2024legalreasoner} (e.g., in predicting court sentencing decisions). 

The performance of a legal LLM can be evaluated based on classification, ranking, and prediction accuracy using traditional evaluation metrics. Examples include precision, recall, accuracy, and F1 score for classification; NDCG and MRR for ranking; and MAE for prediction. Readers are referred to \cite{yacouby2020probabilistic,jarvelin2002cumulated,chapelle2009expected} for the technical definitions of these metrics. 

In addition, there are tasks designed to evaluate legal LLMs for their reliability and safety. For example, {\it Super Legal Bench}\footnote{\url{https://data.court.gov.cn/pages/modelEvaluation.html}} is a large-scale Chinese legal LLM testing set. Some tasks provided in the testing set assess LLMs' responses in terms of national security, public security, ethics, and morality. There are also tasks that evaluate models' speed performance. For example, in~\cite{legal-llm-evaluation-draft}, tasks are designed to evaluate the response time and processing speed of legal LLMs. While these aspects are crucial in production-level deployment, they are generally less related to the core legal reasoning ability of the models.

\subsection{Multi-Task Benchmarks}

A single task cannot fully evaluate the performance of LLMs in real-world legal scenarios. As Table \ref{tab:combined_benchmarks} shows, an increasing number of multi-task legal benchmarks have been proposed to evaluate LLMs. Those benchmarks that include an extensive collection of generative and decision tasks designed to evaluate legal LLMs across multiple capabilities~\cite{fei2023lawbench}. These include the ability to recite legal knowledge, perform legal reasoning, and produce responses that adhere to ethical standards. There are also focus-specific benchmarks that evaluate legal LLMs on specific aspects. For example, some assess legal language understanding ability through tasks like legal text classification and entailment \cite{chalkidis2021lexglue}, while others evaluate a model's ability to generate legally sound responses with appropriate citations \cite{zhang2024citalaw}. Given that logical reasoning is central to legal practice, recent benchmarks have focused more on assessing LLMs' legal reasoning capabilities \cite{legalbench,dai2023laiw} and their robustness against adversarial attacks at different stages of legal reasoning\cite{hu2025jandh}. 

In this section, we provide a detailed review of such multi-task benchmarks.
% We analyze them in terms of their task design, datasets, and metrics, and further examine to what extent they address the challenges identified in legal LLM applications. 
We present three most well-known benchmarks from the perspectives of outcome correctness, legal reasoning, and trustworthiness, and provide detailed analyses, along with a discussion of their strengths and limitations.

\begin{table*}[t]
    \small
    % \scriptsize
    \centering
    \setlength{\tabcolsep}{3pt} 
    
    \newcolumntype{L}{>{\raggedright\arraybackslash}X}
    
    \caption{Comprehensive Overview of Legal Benchmarks} 
    \label{tab:combined_benchmarks}
    
    \begin{tabularx}{\textwidth}{@{} l p{2.8cm} L L p{1.6cm} L l l @{}}
        \toprule
        & & \multicolumn{4}{c}{\textbf{Description}} & & \\
        \cmidrule(lr){3-6}
        \textbf{Type} & \textbf{Name} & \textbf{Main Task} & \textbf{Source} & \textbf{Scale} & \textbf{Metrics} & \textbf{Lang.} & \textbf{Country} \\
        \midrule

        \multirow{12}{*}{\makecell[l]{Multi\\Legal\\Task}} 
        & LexGLUE\cite{chalkidis2021lexglue} & Classification \& Question Answering & 7 existing English legal datasets & \makecell[l]{$\sim$235,000 \\instances} & Micro-F1, Macro-F1 & EN & USA \\ 
        
        & LegalBench\newline\cite{legalbench}& Multi-dimensional Legal Reasoning & 36 distinct legal corpora & \makecell[l]{$\sim$91,206 \\samples} & Accuracy, Balanced Accuracy, F1 & EN & USA \\ 
        
        & LBOX OPEN\newline\cite{hwang2022multi}& Judgment Prediction, Summarization, Classification & Precedents from South Korean & \makecell[l]{$\sim$50,000 \\ samples} & Exact Match, F1, ROUGE & KO & KR \\ 
        
        & LegalLAMA\cite{chalkidis2023lexfiles} & Legal Knowledge Probing & LeXFiles corpus & 8 sub-tasks & Mean Reciprocal Rank, P@1 & EN & Multi. \\ 
        
        & LEXTREME\newline\cite{niklaus2023lextreme} & Classification, NER & 11 datasets from Law-NLP literature & \makecell[l]{$\sim$1 million \\samples} & Macro-F1, LEXTREME Score & ML & Multi. \\ 
        
        & GENTLE\newline\cite{aoyama2023gentle}& Cross-genre Linguistic Evaluation & Wiki, GitHub, Contracts, etc. & 8 genres & Accuracy, Kappa, UAS/LAS, F1 & EN & USA \\ 
        
        & SCALE\newline\cite{rasiah2023scale} & Classification, Retrieval, Summarization, NER & Swiss Federal Supreme \& Cantonal Courts & $\sim$850,000 samples & Macro-F1, ROUGE/BLEU, Recall@k & 5 Lgs. & CH \\ 
        
        & LawBench\newline\cite{fei2023lawbench} & Memorization, Application & Existing Chinese datasets & $\sim$10,000 questions & Accuracy, F1, rc-F1, soft-F1, ROUGE-L & ZH & CN \\ 
        
        & LAiW\newline\cite{dai2023laiw}  & Legal Practice Logic Evaluation & Existing Chinese datasets & \makecell[l]{11,605\\questions} & Accuracy, Miss Rate, F1, Mcc, ROUGE & ZH & CN \\ 
        
        & LexEval\newline\cite{li2024lexeval} & Comprehensive Legal Ability Classification & Existing datasets, bar exams & \makecell[l]{14,150\\questions} & Accuracy (Multiple Choice), ROUGE-L & ZH & CN \\ 

        & LegalAgentBench\newline\cite{li2024legalagentbenchevaluatingllmagents} & \makecell[l]{Multi-hop reasoning, tool \\ utilization, legal writing} & 17 real-world legal corpora & \makecell[l]{300 tasks, \\ 37 tools} & \makecell[l]{Success Rate, BERTScore \\ Progress Rate} & ZH & CN \\
        \bottomrule
    \end{tabularx}
\end{table*}

\begin{enumerate}

\item{\textbf{LexEval}}~\cite{li2024lexeval}

For \textbf{output accuracy}, we select LexEval as our example for analysis. LexEval is currently the most comprehensive benchmark for evaluating the output accuracy of models on legal tasks. It is therefore the most representative choice for detailed analysis. Other benchmarks \cite{fan2025lexam,fei2023lawbench} face future improvement directions that are similar to those of LexEval, and thus can be analyzed and discussed using LexEval as a reference.

\paragraph{Task Design}

LexEval is grounded in real-world legal tasks, reflecting how legal professionals manage, analyze, and resolve legal issues in practice. It proposes a Legal Cognitive Ability Taxonomy that characterizes the capabilities required for LLMs. At the \textbf{Memorization} level, models are assessed on their ability to recall fundamental legal knowledge, such as statutes, case law, and legal terminology. The \textbf{Understanding} dimension examines whether models can correctly interpret the meaning and implications of legal texts and problem descriptions beyond surface-level recall. \textbf{Logical Inference} evaluates the model's capacity for legal reasoning, including deriving conclusions from facts and rules and applying legal principles appropriately. The \textbf{Discrimination} dimension focuses on assessing the relevance and applicability of legal information under specific contexts. At the \textbf{Generation} level, models are expected to produce well-structured, legally sound texts, such as legal documents and arguments, within defined scenarios. Finally, the \textbf{Ethics} dimension evaluates whether models can identify and reason about legal ethical issues while considering professional responsibility and social values.

\begin{table*}[!h]
    \small
    % \scriptsize
    % \centering
    % 调整列间距，使5列布局更紧凑
    \setlength{\tabcolsep}{5pt} 
    \renewcommand{\arraystretch}{0.4}
    % 定义 L 为自动换行的左对齐列
    \newcolumntype{L}{>{\raggedright\arraybackslash}X}
    
    \caption{Comprehensive Overview of Benchmarks} 
    \label{tab:legal_benchmarks_v2}
    
    % 设置总宽度为 \textwidth，Description 列(L) 会自动调整
    \begin{tabularx}{\textwidth}{@{} p{1.1cm} p{4cm} p{10.1cm} L L @{}}
        \toprule
        \textbf{Type} & \textbf{Name} & \textbf{Description} & \textbf{Lang.} & \textbf{Country} \\
        \midrule

        % --- Case Retrieval ---
        \multirow{7}{*}{\makecell[l]{Case\\Retrieval}} 
        & LeCaRD~\scriptsize{\cite{ma2021lecard}}& 107 queries and 10,700 candidates from 43,000 Chinese criminal judgements. & ZH & CN \\ \addlinespace[3pt]
        & LeCaRDv2~\scriptsize{\cite{li2024lecardv2}} & Bigger dataset: 800 query cases \& 55,192 candidates from 4.3M documents. & ZH & CN \\ \addlinespace[3pt]
        & COLIEE~\scriptsize{\cite{kim2022coliee} }& Annual competition for legal IR and statute law entailment across jurisdictions. & EN/JP & CA/JP \\ \addlinespace[3pt]
        & doc. similarity~\scriptsize{\cite{mandal2021legal}} & Similarity scores (0-1) for 53,000 Supreme Court cases and 12,000 Indian Acts. & EN & IN \\ \addlinespace[3pt]
        & MUSER~\scriptsize{\cite{li2023muser}} & Multi-view retrieval for Chinese civil cases covering facts, disputes, and laws. & ZH & CN \\ \addlinespace[3pt]
        & ML2IR~\scriptsize{\cite{phyu2024ml2ir}} & GraphRAG bench for low-resource legal IR using Burmese cases and statutes. & MY & MM \\ \addlinespace[3pt]
        & ELAM\scriptsize{\cite{yu2022explainable}} & 5,000 case pairs for explainable matching with gold-standard rationales. & ZH & CN \\
        \midrule

        % --- Question Answering ---
        \multirow{9}{*}{\makecell[l]{Question\\Answering}} 
        & PrivacyQA {\scriptsize\cite{ravichander2019question}} & 1,750 QA pairs on privacy policies with expert-level annotations. & EN & USA \\ \addlinespace[3pt]
        & SARA{\scriptsize~\cite{holzenberger2020dataset}} & Statutory reasoning dataset based on US tax law and corresponding test cases. & EN & USA \\ \addlinespace[3pt]
        & JEC-QA~\scriptsize{\cite{zhong2020jec}} & Large-scale data of 26,365 questions from the Chinese National Judicial Exam. & ZH & CN \\ \addlinespace[3pt]
        & CaseHOLD~\scriptsize{\cite{zheng2021does}} & 53,000 multiple-choice questions focusing on judicial holdings in US case law. & EN & USA \\ \addlinespace[3pt]
        & LEXAM\scriptsize{\cite{fan2025lexam}} & 4,886 legal questions sourced from law exams at the University of Zurich. & DE/EN & Multi. \\ \addlinespace[3pt]
        & EQUALS\scriptsize{\cite{chen2023equals}} & 6,914 LQA triplets featuring precise evidence span and rationale annotations. & ZH & CN \\ \addlinespace[3pt]
        & LegalCQA\scriptsize{\cite{jiang2024hlegalki}} & Public-professional dialogue dataset harvested from online legal advice forums. & ZH & CN \\ \addlinespace[3pt]
        & LLeQA\scriptsize{\cite{louis2024interpretable}} & 1,868 expert-annotated questions based on Belgian civil and statutory law. & FR & BE \\ \addlinespace[3pt]
        & Legal-LFQA\scriptsize{\cite{ujwal2024reasoning}} & 18,000 instances for long-form QA, structured around the IRAC framework. & EN & USA \\
        \midrule

        % --- Ethics of Legal AI ---
        \multirow{3}{*}{\makecell[l]{AI Ethics}} 
        & J\&H\scriptsize{\cite{hu2025jandh}} & The robustness of LLMs against knowledge-injection in syllogistic reasoning. & ZH & CN \\ \addlinespace[3pt]
        & JudiFair\scriptsize{\cite{hu2025llms}} & 177,000 cases for measuring inconsistency, bias, and imbalanced inaccuracy. & ZH & CN \\ 
        \midrule

        % --- Document Classification ---
        \multirow{6}{*}{\makecell[l]{Document\\Class.}} 
        & CAIL2018~\scriptsize{\cite{xiao2018cail2018} }& 2.6M cases for joint prediction of applicable law articles, charges, and terms. & ZH & CN \\ \addlinespace[3pt]
        & German Rental\scriptsize{~\cite{glaser2018classifying}} & 913 sentences from tenancy law categorized by fine-grained semantic types. & DE & DE \\ \addlinespace[3pt]
        & ECHR~\scriptsize{\cite{chalkidis2019neural} }& 11,500 cases from the European Court of Human Rights for judicial prediction. & EN & EU \\ \addlinespace[3pt]
        & EURLEX57K~\scriptsize{{\newline\cite{chalkidis2019extreme}}} & 57,000 legislative documents annotated with concepts from the EUROVOC. & EN & EU \\ \addlinespace[3pt]
        & SwissJudgm.~\scriptsize{\cite{niklaus2021swiss}} & 85,000 cases from the Swiss Federal Supreme Court. & ML & CH \\ \addlinespace[3pt]
        & Ger. Dec. Corp.~\scriptsize{\cite{urchs2021design}} & 32,000 German court decisions enriched with comprehensive expert metadata. & DE & DE \\
        \midrule
        
        % --- Reasoning & LJP ---
        \multirow{6}{*}{\makecell[l]{Reasoning \\ \& Legal \\Judgment \\ Prediction}} 
        & SLJA~\scriptsize{\cite{deng2023syllogistic}} & 11,000 cases for multi-task reasoning, including case retrieval and LJP. & ZH & CN \\ \addlinespace[3pt]
        & FSCS~\scriptsize{\cite{niklaus2021swiss}} & 85,000 cases for evaluating prediction accuracy and model robustness. & ML & CH \\ \addlinespace[3pt]
        & MultiLJP~\scriptsize{\cite{lyu2023multi}} & 23,000 cases designed for legal reasoning in complex multi-defendant scenarios. & ZH & CN \\ \addlinespace[3pt]
        & MSLR~\scriptsize{\cite{yu2025benchmarking}} & 1,400 insider trading cases utilizing the IRAC framework for legal reasoning. & ZH & CN \\ \addlinespace[3pt]
        & PRBench~\scriptsize{\cite{akyurek2025prbench}} & 1,100 tasks focused on high-stakes professional reasoning for practitioners. & EN & USA \\
        \midrule

        % --- Summarization ---
        \multirow{7}{*}{\makecell[l]{Summariz-\\ation}} 
        & BillSum\scriptsize{\newline\cite{kornilova2019billsum}} & 22,000 Congressional bills paired with high-quality expert-written summaries. & EN & USA \\ \addlinespace[3pt]
        & CLSum~\scriptsize{\cite{liu2024low}} & Summaries of common law judgments from Canada, Australia, the UK, and HK. & EN & Multi. \\ \addlinespace[3pt]
        & Plain Contract~\scriptsize{\cite{manor2019plain}} & 446 parallel pairs of technical contracts and their plain-English simplifications. & EN & USA \\ \addlinespace[3pt]
        & Priv.PolicySum\scriptsize{\cite{wilson2016creation}} & Multi-source privacy policy summaries extracted from 151 global tech-firms. & EN & - \\ \addlinespace[3pt]
        & EUR-Lex-Sum{\scriptsize\newline\cite{aumiller2022eur}} & 1,500 document-summary pairs per language across 24 official EU languages. & ML & EU \\ \addlinespace[3pt]
        & Multi-LexSum{\scriptsize~\cite{shen2022multi}}& 9,280 summaries for multi-juris. case law to test multi-doc summarization. & EN & - \\ \addlinespace[3pt]
        & IN-bs / Ext~\scriptsize{\cite{shukla2022legal}}  & Bench for abstractive and extractive summarization of Indian Supreme Court. & EN & IN \\
        \midrule
        
        % --- Entity Extraction ---
        \multirow{5}{*}{\makecell[l]{Entity\\Extraction}} 
        & Contract Elem.\scriptsize{\newline\cite{chalkidis2017extracting}} & 3,500 English contracts with gold-standard annotations of key contract elements. & EN & UK \\ \addlinespace[3pt]
        & LEVEN\scriptsize{~\cite{yao2022leven}} & A legal event extraction dataset covering 108 distinct event types in Chinese law. & ZH & CN \\ \addlinespace[3pt]
        & CDJUR-BR\scriptsize{~\cite{mauricio2023cdjur}} & 44,000 annotations for 21 fine-grained legal entity types in Brazilian Portuguese. & PT & BR \\ \addlinespace[3pt]
        & InLegalNER~\scriptsize{\cite{kalamkar2022named}} & Domain-specific NER for Indian judgments, petitioners, courts, and judges. & EN & IN \\ \addlinespace[3pt]
        & JointExtraction~\scriptsize{\cite{chen2020joint}} & Joint entity and relation extraction dataset based on drug-related criminal cases. & ZH & CN \\
        \midrule

        % --- Others ---
        \multirow{6}{*}{Others} 
        & VerbCL\scriptsize{~\cite{rossi2021verbcl}} & Large-scale citation graph of court opinions supporting legal arguments. & EN & USA \\ \addlinespace[3pt]
        & ContractNLI\scriptsize{~\newline\cite{koreeda2021contractnli}} & Natural Language Inference for contracts featuring 607 annotated documents. & EN & USA \\ \addlinespace[3pt]
        & FairLex\scriptsize{~\cite{chalkidis2022fairlex}} & Bench for fairness across 4 jurisdictions, 5 languages, and 5 protected attributes. & ML & Multi. \\ \addlinespace[3pt]
        & Demosthen\scriptsize{~\cite{grundler2022detecting}} & Argument mining dataset based on 40 CJEU decisions regarding fiscal state aid. & EN & EU \\ \addlinespace[3pt]
        & MultiLegalSBD\scriptsize{~\newline\cite{brugger2023multilegalsbd}} & Multilingual Sentence Boundary Detection bench with 130k legal sentences. & ML & Multi. \\ \addlinespace[2pt]
        & MAUD\scriptsize{~\cite{wang2023maud}} & 39k comprehension tasks (ABA Public Target Deal Points Study). & EN & USA \\
        \bottomrule

    \end{tabularx}
\end{table*}

\paragraph{Dataset and Metrics}
The benchmark dataset consists of three components: (i) existing legal datasets, including CAIL\cite{xiao2018cail2018}, JEC-QA\cite{zhong2020jec}, and LeCaRD\cite{ma2021lecard}; (ii) questions derived from the National Uniform Legal Profession Qualification Examination; and (iii) an expert-annotated dataset curated specifically for this benchmark. For multiple-choice questions, Accuracy is adopted as the evaluation metric. For generation tasks, model performance is evaluated using ROUGE-L.

\paragraph{Summary}

This benchmark decomposes LLMs' legal capabilities into six distinct dimensions and evaluates each dimension separately, resulting in a structured and comprehensive assessment framework. Each dimension includes a diverse set of tasks, covering both objective and generative questions with a relatively large number of evaluation instances.

However, limitations remain when considering more realistic and fine-grained evaluation in legal practice. Many tasks are highly standardized and fail to reflect real-world legal scenarios. For instance, privacy-related tasks primarily assess knowledge of privacy law rather than a model's ability to protect user privacy in practical legal interactions. The heavy reliance on examination-style questions further restricts the benchmark's ability to capture the complexity of real cases, which often involve ambiguous, redundant, or misleading information. Consequently, the benchmark mainly evaluates legal knowledge and reasoning in idealized settings rather than performance in realistic legal workflows.

Moreover, the evaluation metrics are relatively coarse. For generation tasks, exclusive reliance on ROUGE-L cannot adequately capture fine-grained legal reasoning, logical coherence, or argument quality. 
Finally, since a substantial portion of the dataset is drawn from previously released public sources, potential data contamination may affect the fairness and reliability of the evaluation results.

% ~\cite{fei2023lawbench} assesses legal LLMs across three cognitive levels: memorization (recall of legal knowledge), understanding (comprehension of legal principles), and application (use of legal reasoning to perform legal tasks). It includes 20 diverse tasks, such as legal entity recognition, reading comprehension, and criminal damage calculation, thereby making it more aligned with real-world legal applications compared with multiple-choice-based evaluations. Additionally, it introduces an ``abstention rate'' metric to measure the frequency with which models refuse to respond or fail to understand instructions. Figure~\ref{fig:LawBench-res} in Appendix~\ref{app:benchmark} shows the evaluation results of 51 LLMs on the LawBench benchmark.

\item{\textbf{LegalBench}}~\cite{legalbench} 

LegalBench is one of the earliest and most well-known datasets for evaluating \textbf{legal reasoning}. It provides a fine-grained decomposition of legal reasoning and covers a large amount of real-world legal data. Selecting this benchmark as a representative example allows us to more clearly illustrate how legal reasoning is incorporated into the evaluation process, and the discussion of this benchmark is also applicable to other benchmarks.

\paragraph{Task Design}
LegalBench consists of 162 legal tasks covering six types of legal reasoning. The benchmark includes tasks structured under the IRAC (Issue, Rule, Application, Conclusion) framework. These tasks involve {\it identifying} legal issues, {\it recalling} legal rules, {\it applying} rules to facts, or {\it concluding} an analysis with a legal outcome. LegalBench identifies six types of legal reasoning that LLMs can be evaluated for: issue-spotting, rule-recall,
rule-application, rule-conclusion, interpretation, and rhetorical-understanding. The objective of the benchmark is to evaluate LLMs and identify areas where they can effectively assist legal professionals.

\paragraph{Dataset and Metrics}
LegalBench datasets are drawn from three sources: 1. existing available datasets and corpora, 2. datasets that were previously constructed by legal professionals but never released, including datasets hand-coded by legal scholars as part of prior empirical legal projects, and 3. developed specifically for LegalBench, by the authors of this paper. Overall, tasks are drawn from 36 distinct corpora.
LegalBench adopts a task-specific evaluation strategy. Decision-based tasks, such as rule recall, rule conclusion, interpretation, rhetorical understanding, and issue spotting, are evaluated using standard automatic metrics, including accuracy and F1 score. Generation-based tasks, particularly rule-application, rely on expert-validated rubric-based evaluation that assesses both correctness and the presence of substantive legal analysis.

\paragraph{Summary}
LegalBench evaluates legal reasoning ability by decomposing the reasoning process into six distinct stages and designing targeted tasks for each stage. This work is among the earliest efforts to systematically break down legal reasoning into fine-grained components and to construct an evaluation benchmark accordingly. Also, the benchmark includes a large number of evaluation instances, which contributes to its broad coverage.

Nevertheless, the evaluation metrics adopted in LegalBench remain relatively lightweight for assessing complex legal reasoning, especially in generation-based tasks. Metrics based on surface-level similarity may not fully capture the correctness, depth, or legal soundness of multi-step reasoning processes. Incorporating more fine-grained, rubric-based evaluation criteria could further strengthen the assessment of reasoning quality beyond textual overlap.

In addition, although the authors suggest that the proposed reasoning framework is applicable beyond U.S. legal cases, legal reasoning practices vary across jurisdictions. While IRAC provides a widely used and intuitive structure for legal reasoning, it aligns more naturally with common law systems and may require adaptation when applied to civil law traditions. This observation points to the potential value of developing evaluation benchmarks that can more explicitly account for cross-jurisdictional differences in legal reasoning, enabling more comprehensive assessment of LLMs across diverse legal systems.

\item{\textbf{JudiFair}~\cite{hu2025llms}}

JudiFair is the first large-scale benchmark for evaluating judicial fairness of LLMs. \textbf{Trustworthiness} will become increasingly important in the evaluation of LLMs in legal scenarios. Therefore, we select this benchmark as an illustrative example, with the aim of inspiring further discussion and encouraging more research on LLM trustworthiness.

\paragraph{Task Design}
JudiFair focuses on evaluating the judicial fairness of LLMs. Grounded in extensive theoretical discussions of fairness in law and moral philosophy, this benchmark proposes a comprehensive and systematic framework for assessing judicial fairness. 
% Judicial fairness is examined from two complementary perspectives: substantive factors and procedural factors. Within each perspective, fairness is further analyzed along both demographic and non-demographic dimensions. 
The fairness criteria are defined by legal experts, resulting in a total of 65 labels across four categories, which together provide a fine-grained representation of legally relevant and irrelevant attributes that may influence judicial outcomes.

\paragraph{Dataset and Metrics}

The JudiFair dataset contains 177,100 case instances annotated with 65 labels, derived from 1,100 judicial documents. It is constructed using a counterfactual annotation strategy, where specific labels are systematically altered while all other case facts remain unchanged (e.g., modifying a defendant's demographic attribute from male to female). For each counterfactual variant, the model generates a sentencing-related output, enabling analysis of whether legally irrelevant attributes lead to unjustified differences in judicial decisions and thus introduce unfairness.

The evaluation framework examines LLM behavior from multiple complementary perspectives. \textbf{Inconsistency} is measured by the frequency of decision changes when only a single legally irrelevant label is modified. \textbf{Bias} is assessed through regression analysis with case-level fixed effects to identify systematic influences of specific attributes. \textbf{Imbalanced inaccuracy} compares prediction errors across groups relative to human judgments, revealing disparities that may disadvantage certain populations. Together, these metrics provide a multidimensional and statistically grounded assessment of judicial fairness in LLM-based decision-making.
\vspace{-0.1cm}
\paragraph{Summary}
Most existing evaluations of LLMs in legal tasks focus primarily on output accuracy and reasoning ability, while relatively little attention has been paid to trustworthiness. JudiFair represents the first large-scale effort to systematically evaluate the fairness of LLMs in judicial applications. Its framework is theoretically well-motivated, drawing on legal scholars to define fairness dimensions, and its evaluation is both detailed and methodologically rigorous, supported by a large and carefully constructed dataset.

However, the scope of JudiFair is limited to the Chinese legal system, and judicial fairness constitutes only one aspect of trustworthiness. Other critical dimensions, such as privacy preservation, safety, and toxicity, remain largely unexplored in current benchmarks. As a result, despite advances in fairness evaluation, there is still insufficient empirical evidence to demonstrate that legal professionals can fully trust LLMs to perform real-world legal tasks without substantial human oversight.

\end{enumerate}

% \vspace{-1.2cm}

\section{Future: Towards Better Evaluation of Legal LLMs}

\subsection{Data Perspective}
In summary, existing studies primarily construct evaluation datasets from three sources: legal examinations, case fact descriptions, and expert annotations. While these data sources are generally reliable and well-controlled, they do not fully simulate the complexity of real-world legal problem-solving. As discussed earlier, practical legal scenarios are inherently more complex, containing redundant information, ambiguous facts, and numerous confounding factors. To advance the evaluation of LLMs in legal tasks, future benchmarks should move beyond exam-style settings and bring models closer to real-world legal environments. This requires incorporating more data derived from realistic legal scenarios and evaluating LLMs under conditions that more faithfully reflect actual judicial and legal practice, thereby enabling a more comprehensive and authentic assessment of their capabilities in handling legal tasks.

\subsection{Method Perspective}
Previous research has covered a wide range of high-level legal tasks; however, it has generally exhibited a strong emphasis on evaluating LLMs' legal capabilities while comparatively overlooking their trustworthiness. Most existing benchmarks focus on tasks such as recalling statutory provisions or predicting judicial outcomes. In contrast, critical dimensions related to trustworthiness, including fairness, safety, and hallucination, remain underexplored. Even highly capable models may pose serious risks if they cannot be used in a trustworthy manner. This concern is particularly acute in the legal domain, where model outputs can directly affect individuals' lives and broader social welfare. Ensuring that LLMs behave responsibly and align with legal and ethical expectations is therefore a fundamental requirement for their deployment.

% Future research should place greater emphasis on evaluating the trustworthiness of LLMs in legal scenarios and on formally defining and assessing specific dimensions such as legal hallucination. Addressing these issues is essential for building confidence among legal practitioners and enabling the responsible integration of LLMs into judicial practice.

\subsection{Metrics Perspective}

Existing work can be categorized into decision and generation tasks. To reflect real-world legal practice, evaluation frameworks should primarily focus on generation tasks, with decision tasks playing a supporting role. In practice, legal reasoning is usually expressed through narrative explanations rather than discrete decisions, and purely decision-based tasks fail to capture the complexity of real legal scenarios. Current evaluation of generative tasks\cite{li2024lexeval,fei2023lawbench} largely relies on surface-level similarity metrics such as ROUGE-L and the LLM-as-a-judge paradigm. However, these approaches are insufficient for legal applications. Surface-level metrics cannot capture fine-grained legal distinctions, where minor wording differences may lead to substantially different legal meanings and consequences. Similarly, the LLM-as-a-judge paradigm is problematic, as models that struggle with legal reasoning themselves may not reliably assess the legal soundness of other models’ outputs. Future evaluation should therefore involve legal experts and incorporate fine-grained legal knowledge, for example through expert-designed rubrics that assess reasoning quality beyond final answers.

% \section{Conclusion}

% \newpage

\appendix

\bibliographystyle{named}
\bibliography{ijcai26}

\end{document}